\documentclass[%
 reprint,
superscriptaddress,
 amsmath,amssymb,
 aps,
prl,
floatfix,
]{revtex4-2}

\usepackage{graphicx}
\usepackage{dcolumn}
\usepackage{bm}
\usepackage{pdfpages}

\makeatletter
\AtBeginDocument{\let\LS@rot\@undefined}
\makeatother

\begin{document}

\preprint{APS/123-QED}

\title{Anisotropic exciton-polaritons reveal non-Hermitian topology in van der Waals materials}

\author{D. Chakrabarty}
\author{A. Dhara}
\author{P. Das}
\affiliation{Department of Physics, Indian Institute of Technology, Kharagpur-721302, India}
\author{K. Ghosh}
\author{A. R. Chaudhuri}
\affiliation{Materials Science Centre, Indian Institute of Technology, Kharagpur-721302, India}
\author{S. Dhara}
\email{sajaldhara@phy.iitkgp.ac.in}
\affiliation{Department of Physics, Indian Institute of Technology, Kharagpur-721302, India}

\date{\today}

\begin{abstract}
Topological band theory has been expanded into various domains in applied physics, offering significant potential for future technologies. Recent developments indicate that unique bulk band topology perceived for electrons can be realized in a system of light-matter quasiparticles with reduced crystal symmetry by utilizing tunable light-matter interaction. In this work we realize topologically non-trivial energy band dispersion of exciton-polaritons confined in two-dimensional anisotropic materials inside an optical microcavity, and show the emergence of exceptional points (EPs) due to non-Hermitian topology arising from excitonic dipole oscillators with finite quasi particle lifetime. Fourier-plane imaging reveals two pairs of EPs connected by bulk Fermi arcs for each of the transverse electric and magnetic polarized modes. An anisotropic Lorentz oscillator model captures the exact band dispersion observed in our experiment in two-dimensional momentum space. Our findings establish anisotropic two-dimensional materials as a platform for exploring non-Hermitian topological physics, with implications for polarization-controlled optical technologies.
\end{abstract}

\maketitle

Topological band theories have been widely explored not only in condensed matter but also in optics with artificially fabricated photonic crystals showing Dirac point like features in the photonic band dispersions \cite{haldanePossibleRealizationDirectional2008,luTopologicalPhotonics2014,ozawaTopologicalPhotonics2019,partoNonHermitianTopologicalPhotonics2021, bergholtzExceptionalTopologyNonHermitian2021, liExceptionalPointsNonHermitian2023,karzigTopologicalPolaritons2015,solnyshkovMicrocavityPolaritonsTopological2021,hePolaritonicChernInsulators2023}. However, recent experiments \cite{xuObservationFermiArc2015,zhouObservationBulkFermi2018, suDirectMeasurementNonHermitian2021,zengTailoringTopologicalTransitions2022, jinObservationPerovskiteTopological2024} and theoretical studies \cite{qianQuantumSpinHall2014,koziiNonHermitianTopologicalTheory2024} reveal that a bulk band topology is quite intrinsic to anisotropic polariton systems without the need of any artificially fabricated photonic lattice. Therefore, the less explored anisotropic two-dimensional materials can provide a true counterpart for topological physics in mesoscopic optics. In fact, it was known from early days that propagation of light in bulk anisotropic media manifests counterintuitive phenomena of conical refraction which is an example of the consequence of topological aspects related to Hamilton’s diabolical point singularities in the ray-surface \cite{berryChapter2Conical2007,berryPhysicsNonhermitianDegeneracies2004}. Recently it has been observed that tunable light-matter interaction in anisotropic media offers a rich platform for the realization of non-Hermitian physics in addition to topologically non-trivial energy bands in polaritons \cite{yuen-zhouPlexcitonDiracPoints2016,richterExceptionalPointsAnisotropic2017, gaoContinuousTransitionWeak2018,enomotoDrasticTransitionsExcited2022, opalaNaturalExceptionalPoints2023a, kedzioraNonHermitianPolaritonPhoton2024}. The anisotropic exciton-polariton system has been recently utilized to achieve a  polarization selective zero-threshold Raman laser which may have potential applications in quantum technology \cite{dharaZerothresholdPTsymmetricPolaritonRaman2025}. Therefore, the non-Hermitian topology of microcavity polaritons in condensed matter system can offer a new avenue for polarization-controlled quantum optics for future technologies.

The emergence of non-Hermitian topology and Fermi arcs has been predicted for 1-T' TMDs in the electronic band structure due to finite quasiparticle lifetime \cite{koziiNonHermitianTopologicalTheory2024}. In this work we show that the finite lifetime of anisotropic excitons in 1-T’ few layer ReS\textsubscript{2} which can be introduced as the damping term in a Lorentz oscillator (LO) model can produce bulk band topology with Fermi arcs present in the polariton bands. ReS\textsubscript{2} offers a rich playground for condensed matter and photonics due to its polarization-dependent excitonic properties \cite{aslanLinearlyPolarizedExcitons2016,dharaAdditionalExcitonicFeatures2020}. In fact, anisotropic two-dimensional materials with strongly bound polarized excitons like ReS\textsubscript{2} have emerged as a unique platform for realizing polarization-sensitive optoelectronic devices \cite{liuHighlySensitiveDetection2016, rahmanAdvent2DRhenium2017}, tunable light-matter coupling \cite{chakrabartyInterfacialAnisotropicExcitonpolariton2021}, polarized exciton-polaritons \cite{gognaSelfHybridizedPolarizedPolaritons2020, coriolanoRydbergPolaritonsReS22022} and wavelength-switchable nanophotonics by leveraging its optical biaxiality \cite{chakrabartyAnisotropicDispersionDielectric2021a, ermolaevWanderingPrincipalOptical2024}. The oscillator strength of anisotropic excitons is dependent on their orientation with respect to the incident electric field due to the reduced symmetry of ReS\textsubscript{2}. Thus, introducing few-layer ReS\textsubscript{2} in the antinode of a microcavity creates anisotropic polariton bands that serve as a unique non-Hermitian system with non-trivial topology, wherein interaction strength between the excitons and cavity modes can be tuned freely between strong and weak light-matter coupling regimes by changing sample orientation. 

\begin{figure*}[ht]
\includegraphics[scale=0.94]{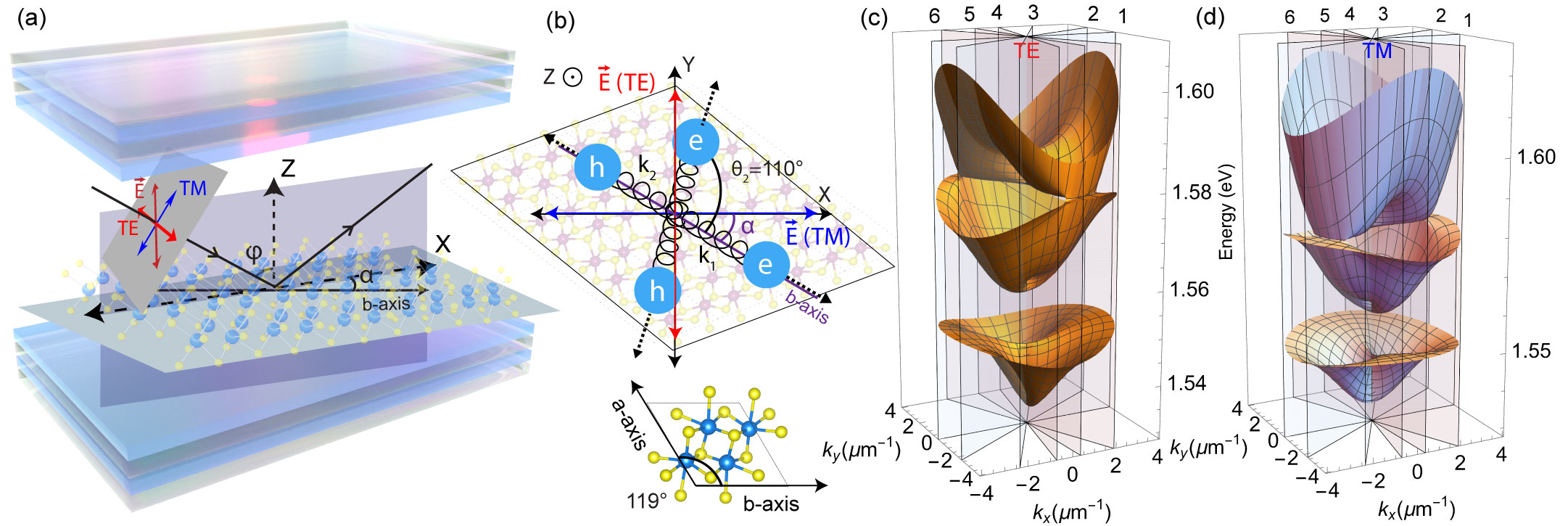}
\caption{\label{fig:1}Realizing anisotropic exciton-polaritons with non-Hermitian topology. (a) Schematic of ReS\textsubscript{2} embedded in a microcavity, showing the TE and TM components of the incident electric field, and the orientation $\alpha$ of the crystallographic b-axis w.r.t. the plane of incidence (X-Z) which is captured by the spectrometer during Fourier imaging. (b) Schematic of the microscopic permittivity model, showing the orientation of the oscillators corresponding to the X\textsubscript{1} and X\textsubscript{2} excitons w.r.t. the crystal b-axis and the incident electric field. (Inset) Unit cell of ReS\textsubscript{2} showing the relative orientation of crystallographic a and b axes. (c-d) Polariton dispersion surfaces in momentum space for TE and TM polarized light obtained from model. Planes labelled 1-6 in the $k_x-k_y$ space represent the specific orientations for which polariton dispersions were probed experimentally.}
\end{figure*}

We experimentally probe the anisotropic polaritonic band structures corresponding to the transverse electric (TE) and transverse magnetic (TM) polarized incident electric field via Fourier plane imaging. The schematic of the microcavity and measurement geometry is shown in Fig \ref{fig:1}(a) (see Methods in Note S1 of the Supplemental Material for details on fabrication and the optical measurement setup). Observing the evolution of polariton band dispersion by changing sample orientation reveals two EP pairs each in momentum space for the TE and TM cases where the polariton branches coalesce, joined by bulk Fermi arcs. We are thus able to realize a topologically non-trivial polaritonic analogue of the bulk Fermi arcs predicted in the band structure of excitonic quasiparticles with finite lifetime. We discuss the origin of the anisotropic polariton band by proposing a model for the complex dielectric permittivity of ReS\textsubscript{2} and obtaining the polariton eigenmodes for the microcavity as discussed below.  

Macroscopically anisotropy is understood as the polarization ($P$) induced by an electric field ($E$), $P_i=\epsilon_0\chi_{ij}E_j$, where $\chi_{ij}$ is the dielectric susceptibility tensor defined in three dimensions. Anisotropy is only exhibited by the thin layer ($\sim$10 nm) of ReS\textsubscript{2} which is sandwiched in the microcavity system made of isotropic dielectric layers. Therefore, near the exciton resonances, the net effect of anisotropic dispersion observed in our experiment is primarily due to the in-plane anisotropy of ReS\textsubscript{2} caused by highly polarized excitonic resonances (X\textsubscript{1} and X\textsubscript{2}) which are unique to this system. A full quantum mechanical understanding of such anisotropic oscillators would involve the details of the orbital symmetries of the valence (metal d orbitals) and conduction (chalcogenide p orbitals) bands in 1-T' ReS\textsubscript{2} similar to earlier works on electronic band structures in 1-T' TMDs \cite{qianQuantumSpinHall2014, koziiNonHermitianTopologicalTheory2024}. We propose a simplified theory of anisotropic Lorentz oscillators (LOs) considering the reduced crystal symmetry resulting in the dipole transitions rates which is expected to have a transition element term proportional to $|\mathbf{\widehat{d}_{1(2)}}.\mathbf{E}|^2$. The two excitons X\textsubscript{1(2)} with different resonance frequencies $\omega_{1(2)}$ can be thought of as independent LOs undergoing damped harmonic oscillations with different spring constants $k_{1(2)}$ in presence of a time varying electric field with frequency $\omega$, as shown in the schematic of Fig. 1(b). The dipole moment of the polarized excitons species X\textsubscript{1(2)} in bare ReS\textsubscript{2} is oriented at $\theta_1=0^{\circ}$ ($\theta_2=110^{\circ}$) with the b-axis \cite{dharaAdditionalExcitonicFeatures2020}, which we denote as the $\widehat{d}_{1(2)}$ directions. The anisotropy in the semiclassical model can be incorporated by constraining the oscillations of LOs in the anisotropic directions {$\widehat{d}_{1(2)}$}. The polarization response $P$ induced along any applied in-plane electric field $E$ due to one of the anisotropic LOs would therefore, depend on the angle ($\theta$) between the electric field and the anisotropic direction of oscillation, which can be obtained as: $P\left(\theta\right)=\epsilon_0\chi_{1(2)}\left(\theta\right)E=\epsilon_0\frac{\omega_{p1(2)}^2\cos^2{(\theta)}}{\omega_{1(2)}^2-\omega^2+i\gamma_{1(2)}\omega}E$, where the squared cosine term appears naturally (see Note S2 for the calculation of the $P\left(\theta\right)$ component from the tensorial form of polarization), signifying the polarization dependent light-matter interaction which is the key feature responsible for the non-trivial topology of polariton dispersion observed in this work. The finite lifetime of exciton, or the damping rate $\gamma_{1(2)}$ is introducing the non-Hermiticity, leading to the formation of Fermi-arcs in the polariton bands.

\begin{figure*}
	\includegraphics[scale=1]{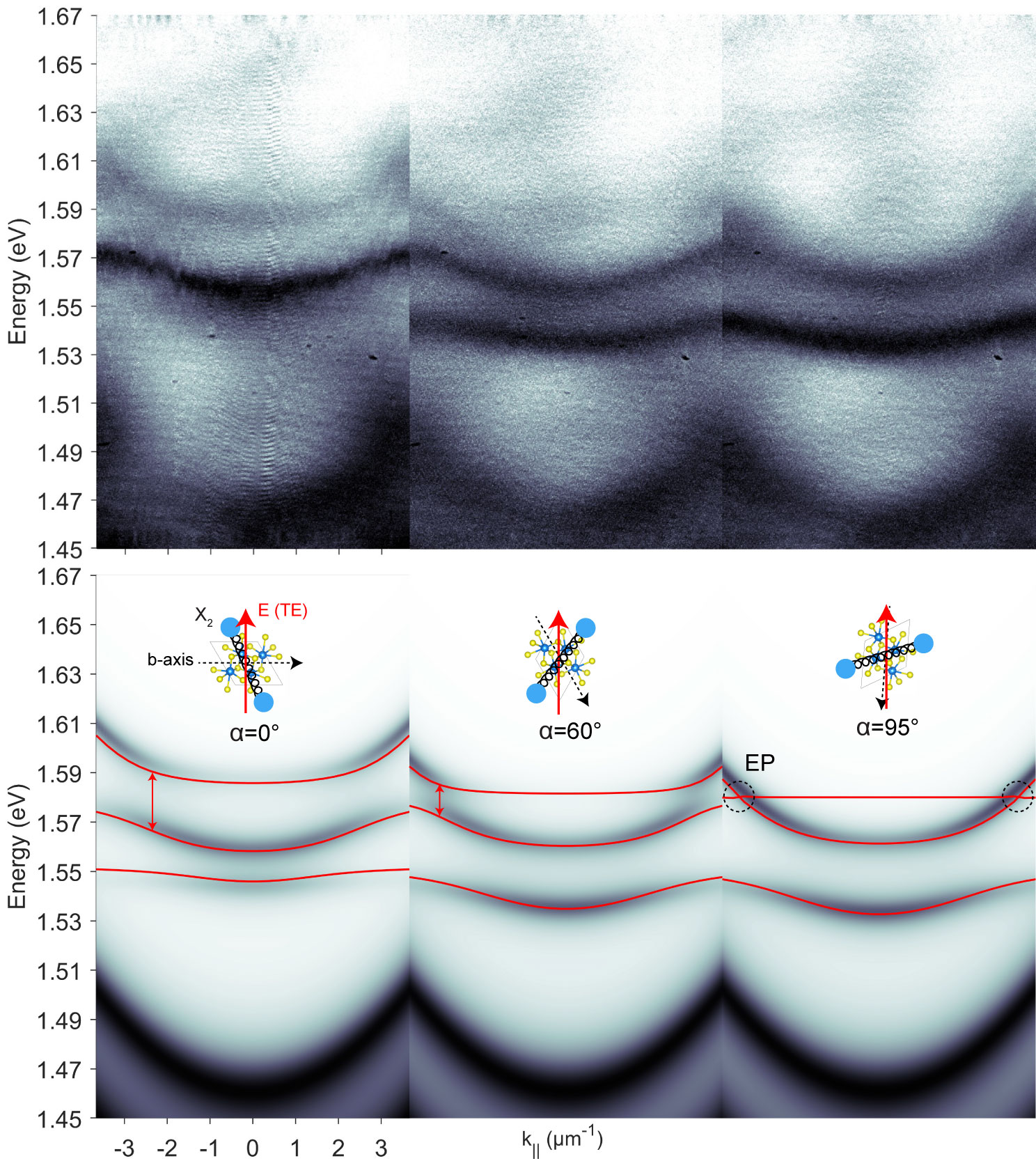}
	\caption{\label{fig:2}Experimental angle-resolved reflectance (upper half) and transfer matrix simulation (lower half) for TE polarized light, corresponding to three different sample orientation angles $\alpha$ as indicated schematically by the middle insets. Polariton modes obtained from theory are superimposed on the simulated reflectance as red lines. Black dotted circles indicate where the UPB and MPB appear degenerate, indicating presence of an EP.} 
\end{figure*}

Using our model of anisotropic oscillators in bare ReS\textsubscript{2} we can now understand the polariton dispersions resulting from strong light-matter interaction inside the cavity. We consider two polarization modes for a given propagation direction of incident light beam, where $\phi$ is the incidence angle and X-Z plane is the plane of incidence: polarization in the plane of incidence (TM) and transverse to the plane of incidence (TE) as shown in Fig 1(a,b) with blue and red arrows respectively.  The angle  between $\widehat{d}_{1(2)}$ and the TE and TM mode electric field is $\theta_{TE}=\frac{\pi}{2}-(\theta_{1\left(2\right)}-\alpha)$ and $\theta_{TM}=(\theta_{1\left(2\right)}-\alpha)$ respectively, where $\alpha$ is the orientation angle between the crystal b-axis and laboratory X-axis. Therefore, by considering the component of electric fields in the above oscillator model, we can express the $\omega$ and $\alpha$ dependent permittivity component along the applied electric field as follows, by considering sum of the contributions from two excitons X\textsubscript{1(2)} and the bulk permittivity: 
\begin{equation} \tag{1a}
\epsilon_{TE}\left(\alpha\right)=\epsilon^b_{TE}(\alpha)+\epsilon_0\left[\chi_1\left(\theta_{TE}\right)+\chi_2\left(\theta_{TE}\right)\right]
\end{equation}
Similarly, we obtain TM mode permittivity by considering the in-plane component of the TM-polarized electric field is proportional to cos($\phi$) (see Note S2):
\begin{multline} \tag{1b}
\epsilon_{TM}\left(\alpha,\phi\right)\ =\epsilon^b_{TM}(\alpha,\phi)+\epsilon_0[\chi_1\left(\theta_{TM}\right)+\\
 \chi_2\left(\theta_{TM}\right)]\cos{\left(\phi\right)} 
\end{multline}
\setcounter{equation}{1}
$\epsilon_{TE(TM)}^b$ are the bulk anisotropy contributions from the material’s principal permittivity tensor away from the excitonic resonance frequencies (see Supplemental Note 3 for the complete mathematical expression). It has been observed the X\textsubscript{1} exciton oscillator has an isotropic component  which is independent of $\alpha$. Thus, we modify the expression of susceptibility to be  $\chi_1\left(\theta\right)=\frac{\omega_{p1}^2\cos^2{(\theta)}{+\omega}_{p1}^{\prime2}}{\omega_{1(2)}^2-\omega^2+i\gamma_{1(2)}\omega}$ where ${\omega\prime}_{p1}^2$ represents the isotropic component of the X\textsubscript{1} exciton oscillator strength. This inclusion is necessary to explain the experimentally observed polariton dispersions, which will be discussed subsequently.

The polariton dispersion for the TE and TM modes can be obtained considering the planar confinement of the cavity mode from the usual relationship of phase velocity and angular frequency \cite{haugQuantumTheoryOptical2009,kavokinMicrocavities2017}:    

\begin{equation}
\omega_{TE(TM)}=\frac{c}{\sqrt{\epsilon_{TE(TM)}}}\sqrt{\frac{m^2\pi^2}{L_{TE(TM)}^2}+k_{\parallel}^2}   
\end{equation}

\begin{figure*}
	\includegraphics[scale=0.94]{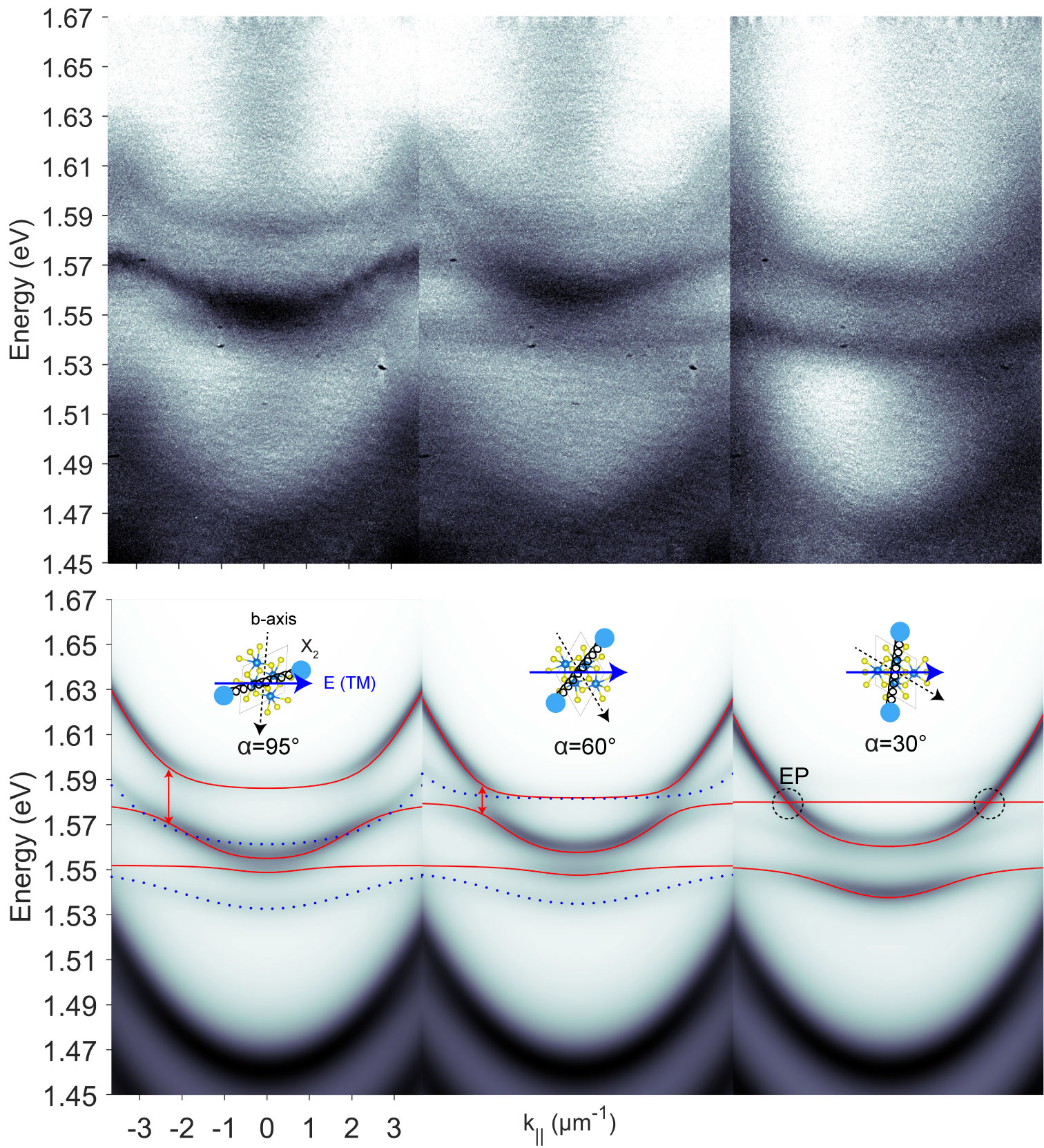}
	\caption{\label{fig:3}Experimental angle-resolved reflectance and simulation revealing EPs in anisotropic polariton dispersion for TM polarized light. Same as Fig. 2, for the case of TM polarized incident light. Blue dotted lines indicate the TE mode polariton branches from the theoretical model which are visible due to polarization mixing.} 
\end{figure*}

Here $L_{TE(TM)}$ is the corresponding cavity length for TE (TM) mode. We set the integer factor m=1 in this case since we use a $\lambda/2$ microcavity in this work, with the thickness of the microcavity SiO\textsubscript{2} layer chosen so that there is only one cavity mode. From Eq. 1(a,b) and Eq. 2, we obtain three polariton branches each in for case of TE and TM polarizations, as a function of sample orientation $\alpha$ and in-plane momentum $k_{\parallel}$ (note that the TM eigenmodes also have an incident angle $\phi$\ dependence). To visualize the polariton bands in the in-plane momentum parameter space, we perform a co-ordinate transformation from $k_{\parallel}-\alpha$ space to $k_x-k_y$ space using the relations $k_x= k_{\parallel}\cos\left(\alpha\right),k_y=k_{\parallel}\sin(\alpha)$. The dispersion of the upper and middle polariton branches (UPB and MPB) for the TE and TM cases, obtained by fitting the theoretical model with experimental results, are shown in Fig 1(c,d). Two pairs of EPs where the two branches coalesce are revealed in the 2D-momentum parameter space for both TE and TM polarizations. Each EP pair is joined by an equi-energy contour known as bulk Fermi arc where the polariton modes are degenerate \cite{bergholtzExceptionalTopologyNonHermitian2021}. In this system, the orientation-dependent light matter coupling between the X\textsubscript{2} exciton and the TE and TM cavity modes, as governed by the anisotropic permittivity model given as Eq. 1(a,b), leads to the formation of EPs in the polariton band structure when the system transitions from the strong to weak coupling regimes and vice-versa.

\begin{figure*}
	\includegraphics[scale=1]{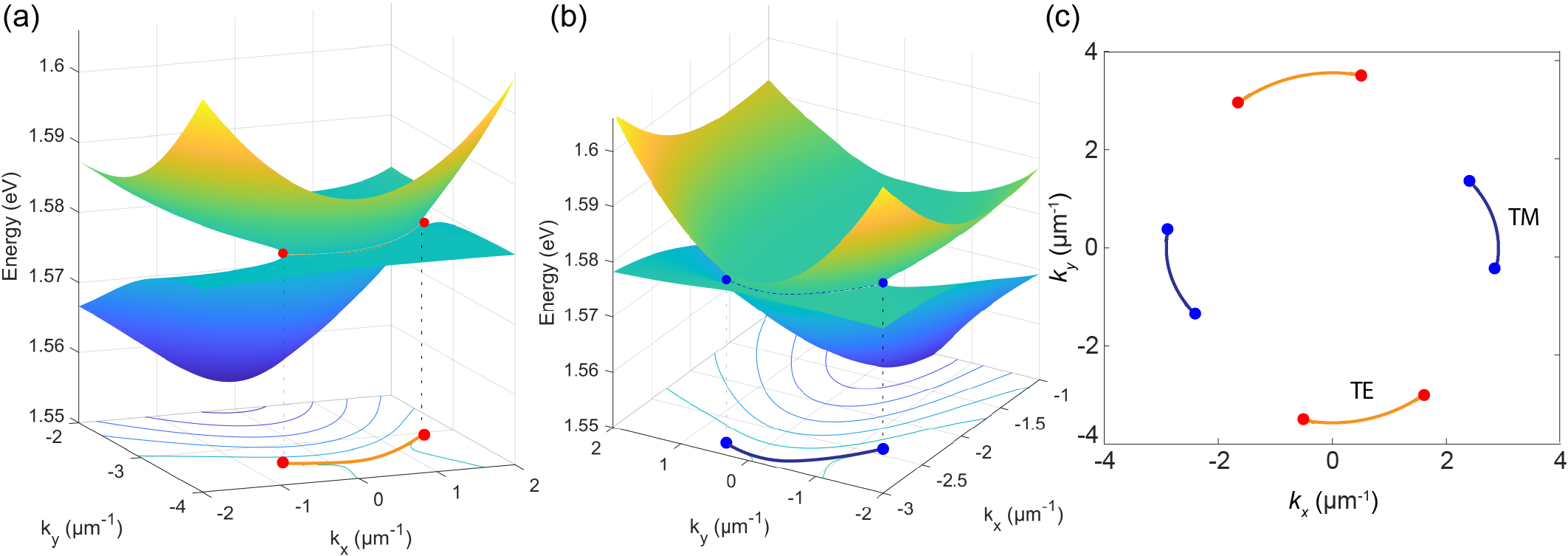}
	\caption{\label{fig:4}Zoomed in view of the UPB and MPB dispersions in $k_x-k_y$ space, revealing the position of the bulk Fermi arcs for (a) TE and (b) TM polarization. Contour plots provided in the $k_x-k_y$ plane represent the energy of the MPB, with the contour lines corresponding to the Fermi arcs marked with orange and blue lines. (c) The equi-energy contours in $k_x-k_y$ space for energy 1.58 eV, indicating the location of the Fermi arcs and EPs.}
\end{figure*}

To reveal the non-Hermitian topology of the polaritonic band structure, we perform angle-resolved reflectance measurements for both TE and TM polarized incident light with the sample mounted in six different orientations. The results for three representative orientations are presented in Figs \ref{fig:2} and \ref{fig:3}, where we observe up to three polariton branches evolving with $\alpha$, corresponding to the three polariton modes we theoretically obtained earlier. By using the plasma frequencies, linewidths and resonance frequencies ($\omega_{p(1,2)}^{2\ }$, $\gamma_{(1,2)}$ and $\omega_{(1,2)})$ of the anisotropic LOs from Eq. 1 and the cavity length  $L_{TE\left(TM\right)}$  from Eq. 2 as fitting parameters, we simultaneously fit the theoretical model with the six sets of polariton dispersions observed for the different orientations (see Table S1 in the Supplemental Material for the parameter values obtained from the fitting, and Figs. S1-S2 for experimental reflectance and simulation for all six measured orientations, along with the position of the uncoupled oscillators and cavity modes). The initial values of $\omega_{p\left(1,2\right)}$, $\gamma_{(1,2)}$ and $\omega_{(1,2)}$ were set based on values obtained from bare ReS\textsubscript{2} of similar thickness (see Figure S1, where X\textsubscript{1} and X\textsubscript{2} resonance energies are 1.538 eV and 1.57 eV). However, since their exact values can depend on the thickness of the ReS\textsubscript{2} sample and the dielectric environment of the microcavity, they were set as fit parameters to account for the slight variation and obtain the best possible fit. The resulting theoretical polariton branches, plotted as red lines in Figs \ref{fig:2} and \ref{fig:3}, show excellent agreement with the experimental data. The experimentally observed reflectance is also well-reproduced by using the 4x4 transfer matrix method (see lower halves of Figs. 2 and 3 for the simulated reflectance) for anisotropic materials as developed by Berreman \cite{berremanOpticsStratifiedAnisotropic1972, passlerGeneralized442017}. The maximum observed Rabi splitting between the UPB and MPB in both TE and TM configurations is $\sim$22 meV. We find the signature of EPs and bulk Fermi arcs in the polariton band, as discussed in detail below.

We can observe the extent of the Fermi arcs and the exact positions of the EPs in parameter space by plotting the entire polariton band structure obtained from the model in $k_x-k_y$ space, which is shown in Fig 1(c) and 1(d) for the TE and TM modes respectively. Measuring the reflectance for each sample orientation $\alpha$ corresponds to probing different planar cross-sections of the polariton band structure in momentum space. These planar cross-sections are labelled 1-6, corresponding to the orientation angles $\alpha = 0^{\circ}, 30^{\circ}, 60^{\circ}, 80^{\circ}, 95^{\circ}$ and $115^{\circ}$. For TE (TM)-polarized incident field, we find the upper and middle polariton branches coalesce between planes 4 and 5 (1 and 2), indicating the presence of an EP in between the corresponding orientations $\alpha$ = $80^{\circ}$ and $95^{\circ}$ ($0^{\circ}$ and $30^{\circ}$) at $k_{\parallel}= \pm3.16\ \mu m^{-1}\ (\pm2.61\ \mu m^{-1})$ where the system transitions from strong to weak coupling regime. The black dotted circles in Fig 2 (Fig 3) mark the region in the experimental data where we find the UBP and MPB are degenerate for TE (TM) mode, indicating the system is in the weak coupling regime. The reflectance data with the theoretical polariton dispersions overlaid, alongside the simulated reflectance is presented in Supplemental Figs S2 and S3 for further clarity. The switching from strong to weak coupling regime is governed by the dependence of the X\textsubscript{2} oscillator strength on $\alpha$ $(\omega_{p2}^2{\sin}^2\left(\theta_2-\alpha\right)$ for TE and $\omega_{p2}^2\cos^2\left(\theta_2-\alpha\right)\cos^2\left(\phi\right)$ for TM mode), as explained by the permittivity model given in Eq. 1(a,b). The coupling strength between the exciton and cavity mode is proportional to the square root of the exciton oscillator strength \cite{andreaniStrongcouplingRegimeQuantum1999, chakrabartyInterfacialAnisotropicExcitonpolariton2021}. When the square root of the oscillator strength becomes less than the average decay rate of the exciton and cavity mode by changing $\alpha$, the system transitions to the weak coupling regime and an EP is realized. Similarly, strong coupling regime is restored between planes 6 and 1 (2 and 3) for TE (TM) mode corresponding to $\alpha=115°$ and 0° (30° and 60°), leading to the formation of a second EP, joined to the previously mentioned EP by a Fermi arc in $k_x-k_y$  space where the UPB and MPB  remain degenerate in energy (1.58 eV). Observing the finer orientation dependence of the polariton dispersions from the theoretical model and transfer matrix simulations reveals the exact orientations where the EPs are realized: $\alpha$=$95^{\circ}$ and $120^{\circ}$ ($5^{\circ}$ and $30^{\circ}$) for TE (TM) mode. Thus, in all $k_x-k_y$ space, four pairs of EPs are realized in this system, two each in the case of TE and TM modes, making this a unique platform for exploring multiple EPs in a tunable non-Hermitian system. Notably, we do not see similar coalescing for the MPB and LPB (see Fig S4 for reflectance cross-section data). This is due to the X\textsubscript{1} exciton having an isotropic component of the oscillator strength which is strong enough to maintain the strong coupling regime for all $\alpha$.

Varying the incident polarization instead of sample orientation can also potentially reveal the topological band structure, but the analysis is complicated due to mixing between the TE and TM modes which are no longer eigenmodes of the microcavity. We can observe some of the TE mode branches for orthogonally polarized TM probe beam in the experimental reflectance. This is due to polarization mixing causing the probe beam to not be perfectly TE or TM polarized, and can be reproduced using the transfer matrix simulation. 

Figure \ref{fig:4} offers a zoomed-in view of the bulk Fermi arcs revealed in the UPB and MPB dispersions corresponding to TE and TM polarization. The EP pairs shown as red (blue) dots for the TE (TM) case are each centered around the point in $k_x-k_y$ space corresponding to sample orientation $\alpha$ for which the coupling strength between the two polariton branches falls to zero [$110^{\circ}$ ($20^{\circ}$) for TE (TM)], governed by the dependence of the oscillator strength of the X\textsubscript{2} exciton on $\alpha$ as discussed previously. The extent of the Fermi arcs, shown as orange (blue) curves joining the EPs in Fig. \ref{fig:4}(a-c) for the TE (TM) case), is dependent on the X\textsubscript{2} exciton and the microcavity photon lifetimes. We observe bulk Fermi arcs spanning $\sim$$25^{\circ}$, corresponding to exciton linewidth of $\sim$6 meV. In the limiting case where there is no loss (i.e., the imaginary component of the complex permittivity is zero), the bulk Fermi arc would vanish and the two EPs would converge to one diabolical point.

\begin{figure}
\includegraphics[scale=0.945]{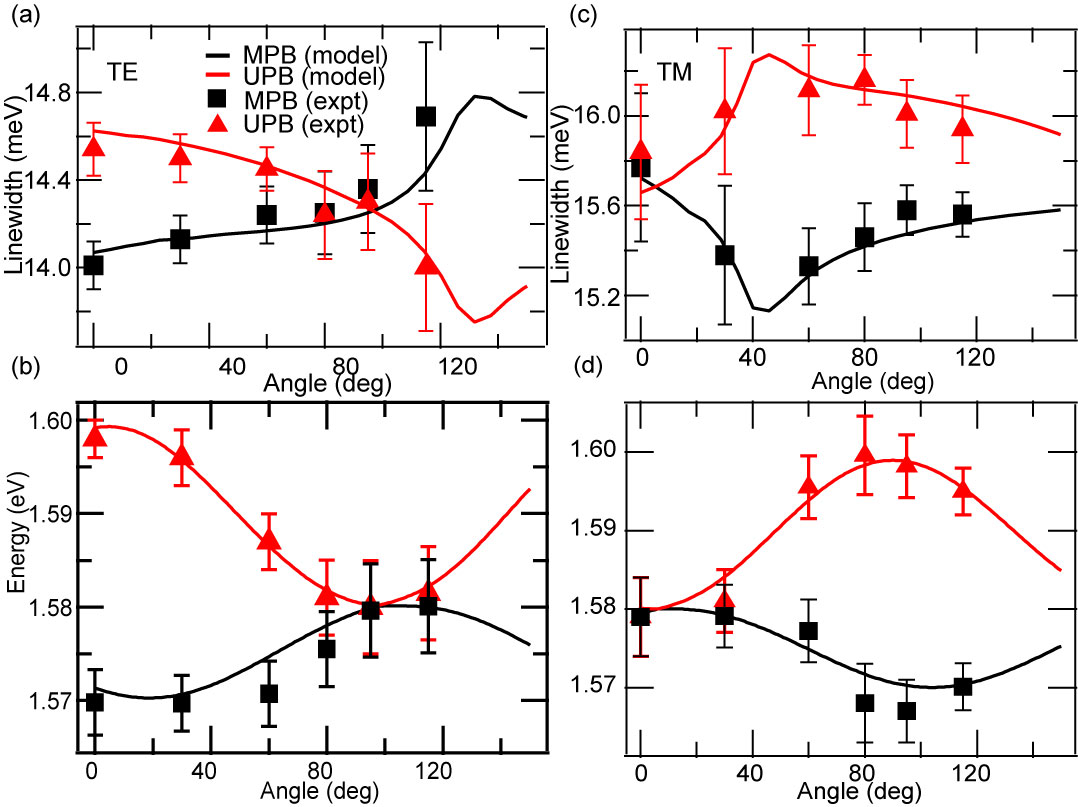}
\caption{\label{fig:5}Extracted and theoretical linewidths (FWHM) and energies of the (a,b) TE and (c,d) TM upper and middle polariton branches for the $k_{\parallel}$ where one of the EPs is observed. Error bars indicate the fitting error.}
\end{figure} 

To look for an additional signature of the EPs, we examine the complex eigenvalues of the polaritons. We extract the linewidths and energies of the polariton branches for TE and TM incidence from the experimental reflectance data by using Lorentzian peak fitting (see Fig. S5 for reflectance cross sections corresponding to the orientations shown in Figs. 2 and 3 and the MPB and UPB peak fitting), the results of which are plotted in Fig \ref{fig:5}(a,b) and (c,d) for TE and TM modes respectively. The square (triangular) markers  indicate the linewidths and energies extracted from the experimental data for the UPB (MPB) at $k_{\parallel}=$ $\sim$$3.1\ \mu m^{-1}$ and $2.6\ \mu m^{-1}$ for the TE and TM modes respectively, in proximity to the Fermi arc where the real part of the polariton eigenvalues are degenerate as shown in Fig \ref{fig:4}(c). We observe that for $\alpha=95^\circ\ (5^\circ)$ for TE (TM) mode the linewidths and energies show a crossing, clearly revealing an EP where the complex eigenvalues of both modes are degenerate. This is followed by a level-repulsion in the imaginary part of the eigenvalues which is characteristic of the Fermi arc where the real values are degenerate but imaginary part is non-degenerate. Since the Fermi arc does not fall inside a semicircle in momentum space due to the changing dispersion of the cavity mode with sample orientation $\alpha$, we obtain a cross section of the arc from these plots at one of the EPs. The second EP can also be seen experimentally, and is shown occurring at $\alpha=30°$ and $k_{\parallel}=$ =$\sim$$2.35\ \mu m^{-1}$ for TM mode in Supplementary Fig S5(b). In addition to the EP shown in Fig 5(a, b) for TE mode, we show evidence the other three EPs occurring at $k_{\parallel}=$ $-$3.16 $\mu m^{-1}$ and $k_{\parallel}=$ $\pm$3.24 $\mu m^{-1}$ in Supplementary Fig. S6. The experimentally observed energies and linewidths are in good agreement with the theoretically obtained dependence on $\alpha$ (plotted as red and black solid lines).

Future work can investigate non-Hermitian topological phases which can be found in the photoluminescence (PL) from the polaritons near the EPs to calculate the associated topological invariants \cite{gianfrateMeasurementQuantumGeometric2020,solnyshkovQuantumMetricWave2021}. However this is complicated in this system due to the quasi-indirect bandgap of ReS\textsubscript{2} making the PL quantum yield significantly less compared to Group VI transition metal dichalcogenides. Further, in angle-resolved photoluminescence measurements only the LPBs were observed (see Supplementary Figure S7). 

In summary, the anisotropic optical response of excitons with finite lifetime in a 1-T` semiconductor was leveraged to realize a polaritonic bandstructure with non-Hermitian topological features, offering new paradigms in polarization-based photonics and EP-based sensors. These topological polaritons can serve as a platform for enabling PT-symmetric lasers  [30] or nanoscale optical devices for polarization and coherence control. We believe that our observation of the Fermi arcs in the polariton dispersion establishes an important equivalence between topological energy bands for electrons and light-matter quasi particles, with both originating from the intrinsic material symmetry.

\begin{acknowledgments}
SD conceptualized the project. DC, AD, KG, ARC and SD contributed in the sample fabrication. DC, AD, PD and SD formulated the experiments, performed data analysis, and developed the theory. DC, AD, performed all optical measurements. DC, AD, and SD drafted the paper, and all authors contributed to reviewing and editing the final draft. SD supervised the project. This work has been supported by funding from the Science and Engineering Research Board (CRG/2018/002845, CRG/2021/000811); Ministry of Education (MoE/STARS- 1/647); Council of Scientific and Industrial Research, India (09/081(1352)/2019-EMR-I); Department of Science and Technology, Ministry of Science and Technology, India (IF180046) and Indian Institute of Technology Kharagpur.
\end{acknowledgments} 

\bibliography{res2mc}

\clearpage


\section*{Supplementary material}

# Anisotropic exciton-polariton dispersion reveals topological Fermi arcs

Devarshi Chakrabarty[1†], Avijit Dhara[1†], Pritam Das[1], Kritika Ghosh[2], Ayan Roy Chaudhuri[2], Sajal Dhara[1*]

[1]*Department of Physics, IIT Kharagpur, Kharagpur-721302, India*

[2]*Materials Science Centre, IIT Kharagpur, Kharagpur-721302, India*

[†] *These authors contributed equally*
[*] *Corresponding author. email: sajaldhara@phy.iitkgp.ac.in*

**List of Contents:**

**Supplementary Note S1-S2**

**Table S1**

**Figs. S1 to S5**

## Note S1: Methods

**Device fabrication**

The microcavity consists of mechanically exfoliated 10 nm ReS$_2$ multilayer embedded between two distributed Bragg reflectors (DBRs) using dry-transfer technique. The bottom mirror consists of 10 pairs of SiO$_2$/Ta$_2$O$_5$, while the top mirror has 8 pairs to allow for preferential out-coupling of light from the cavity in the direction of collection. The center wavelength $\lambda_{DBR}$ of the DBRs is approximately 735 nm. Optical microscope image of the sample is given as Fig S5 in the Supplemental Material.

**Optical measurements**

All measurements were taken at 4K, with the sample mounted in a closed-cycle Helium cryostat. To perform incident polarization-resolved reflectance, white light from a broadband halogen source is passed through a polarizer followed by a half-wave plate, which is rotated to change the angle of incident polarization $\theta$. A 60x objective lens with 0.7 NA is used to focus the beam onto the sample. The fixed spectrometer slit captures a particular plane of incidence and is oriented such TM (TE) polarized reflectance is measured when incident beam polarization $\theta = 0°$ (90°). Vignetting due to the limited aperture of the tube lens collecting the reflected light, causes the effective observation of dispersion up to reflectance angle $\phi \sim 26°$.

## Note S2: Bulk permittivity calculation from bare cavity dispersion

The component of the bulk permittivity for the TE and TM modes are given as follows:

$$\epsilon_{TE}^b(\alpha) = \epsilon_x \sin^2(\alpha) + \epsilon_y \cos^2(\alpha)$$

$$\epsilon_{TM}^b(\alpha, \phi) = [\epsilon_x \cos^2(\alpha) + \epsilon_y \sin^2(\alpha)] \cos^2(\phi) + \epsilon_z \sin^2(\phi)$$

$\epsilon_{(x,y,z)}$ in the above equations are the bulk anisotropy contributions from the material's principal permittivity tensor away from the excitonic resonance frequencies. The values of $\epsilon_x$ and $\epsilon_y$ are obtained using the mean-approximation method [1] for the composite SiO$_2$-ReS$_2$ cavity, where the complex refractive index of ReS$_2$ was taken from previous studies [2,3]. Thickness of the ReS$_2$ layer is $L_{ReS2}$ = 13.4 nm, and the SiO$_2$ cavity length is $\frac{\lambda_{cav}}{2n_{SiO2}}$, where $\lambda_{cav}$ = 735 nm and $n_{SiO2}$ = 1.47. $\epsilon_z$ is approximated to be the same as the dielectric constant of SiO$_2$.

The cavity lengths used in Eqn (2), $L_{TE}$ = 232.5 nm, $L_{TE}$ = 204.8 nm is obtained by fitting the room temperature cavity dispersion as shown in Fig S4, while assuming the exciton oscillator strengths to be zero.

**Table S1: Parameter values used in semiclassical Lorentz oscillator model**

| Oscillator | Orientation $\theta_i$ wrt b-axis (deg) | Plasma frequency (eV) | Resonance frequency (eV) | Linewidth (meV) | Isotropic component (eV) |
|---|---|---|---|---|---|
| X$_1$ | 0 | 0.34 | 1.548 | 6.0 | 0.18 |
| X$_2$ | 110 | 0.32 | 1.58 | 5.7 | 0 |

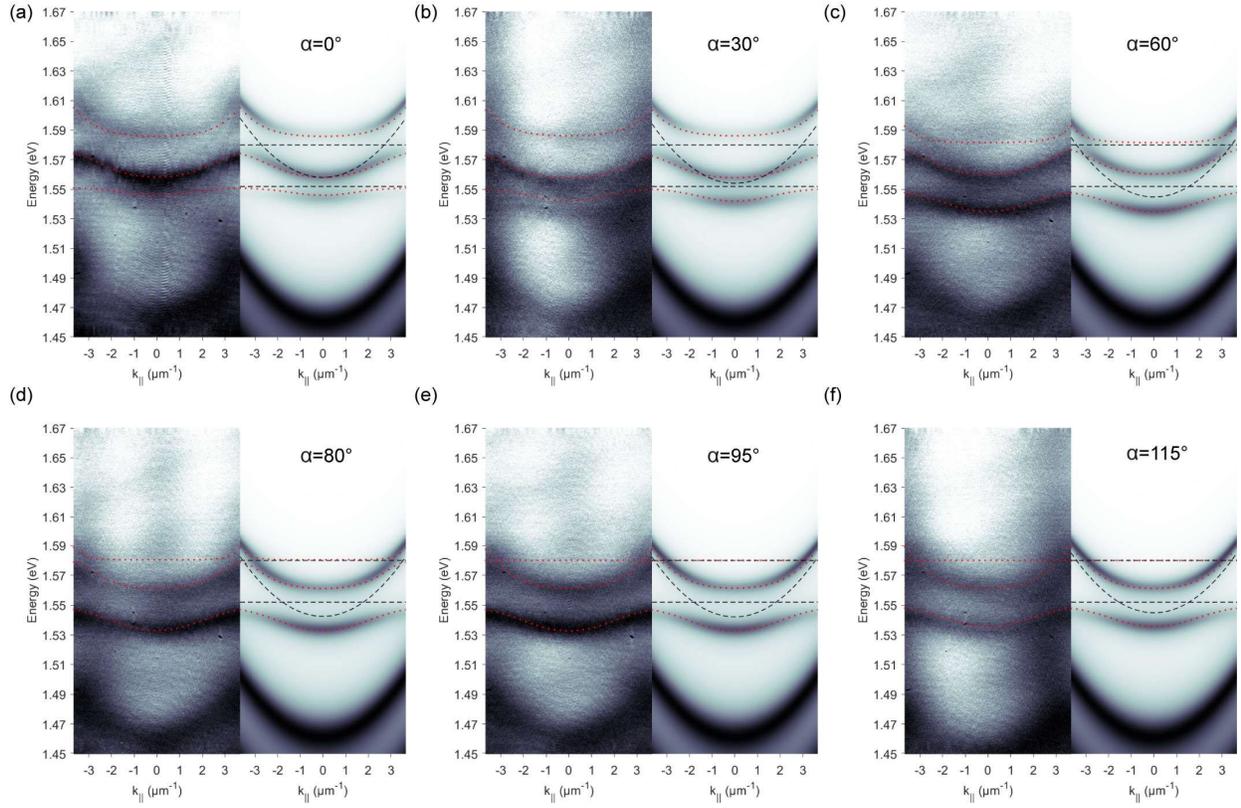

**Fig S1.** Experimental angle resolved reflectance and transfer matrix simulation for TE-polarized incident light for different sample orientations $\alpha$. Red dotted indicate the polariton modes obtained from theory, and black dashed lines indicate the uncoupled exciton and cavity dispersions.

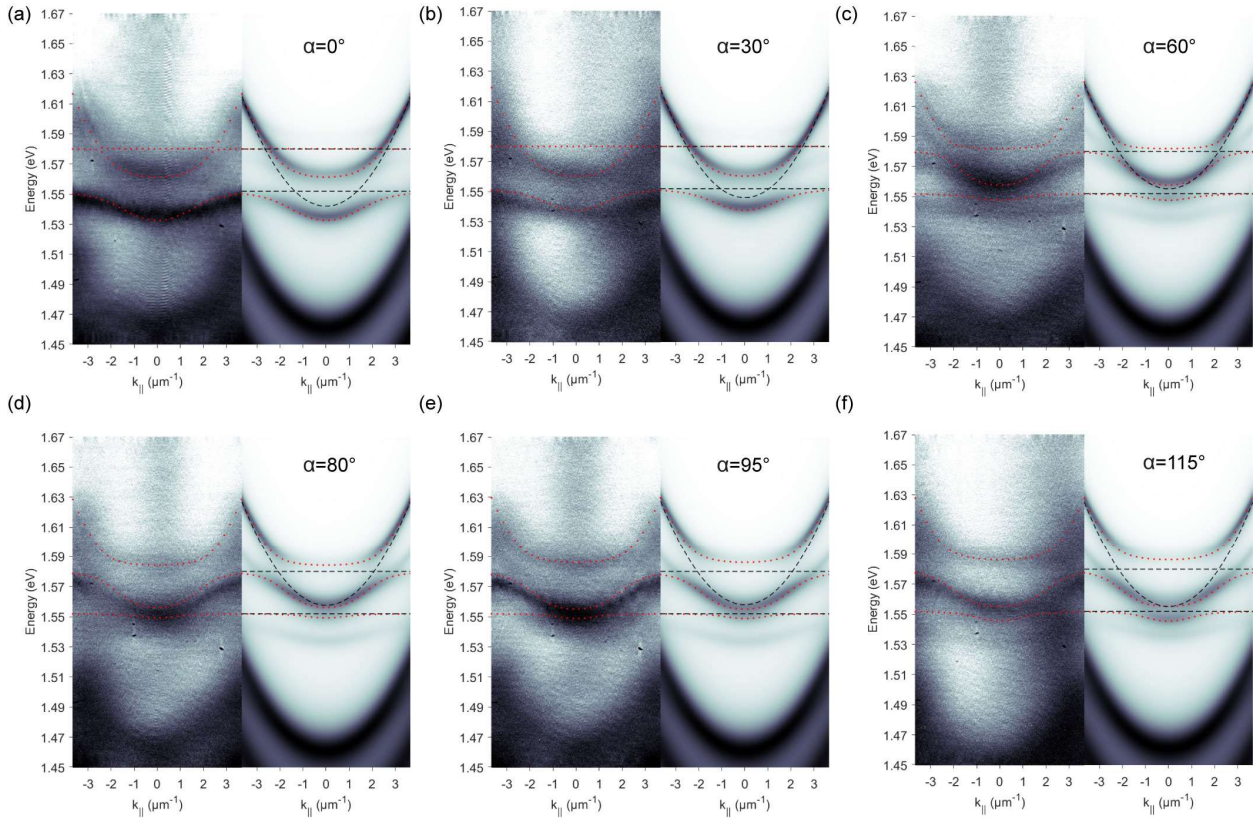

**Fig S2.** Experimental angle resolved reflectance and transfer matrix simulation for TM-polarized incident light for different sample orientations $\alpha$. Red dotted lines indicate the polariton modes obtained from theory, and black dashed lines indicate the uncoupled exciton and cavity dispersions.

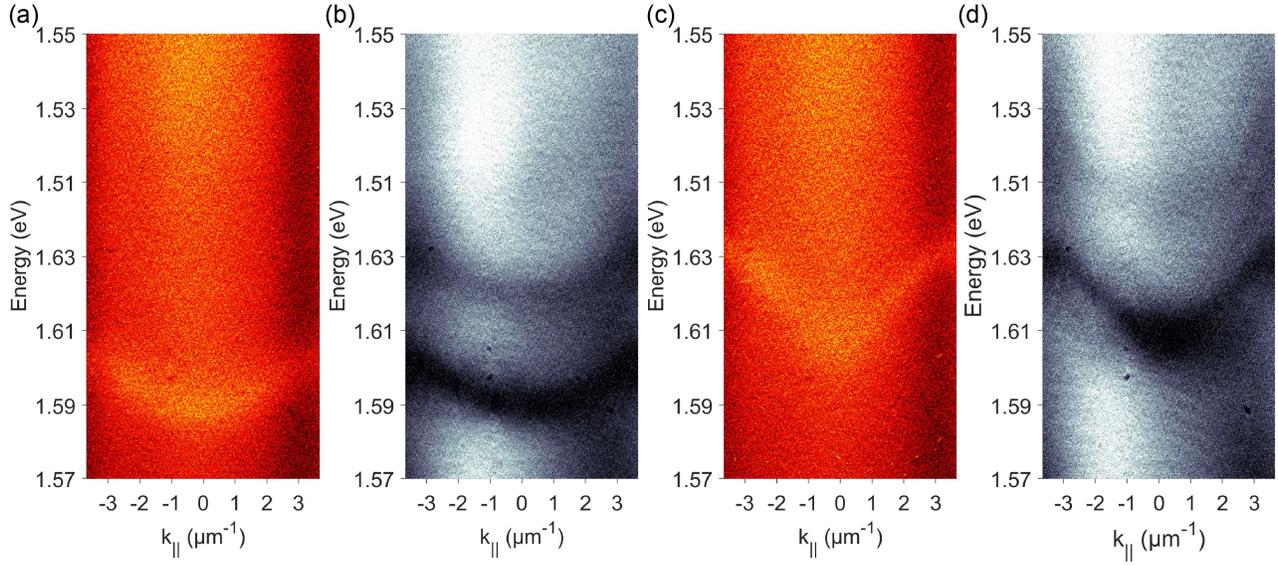

**Fig S3. Angle-resolved photoluminescence for sample orientation α = 0°.** (a, b) PL emission from the LPB, shown alongside angle-resolved reflectance where both MPB and LPB are visible. (c, d) Same as (a, b) but for

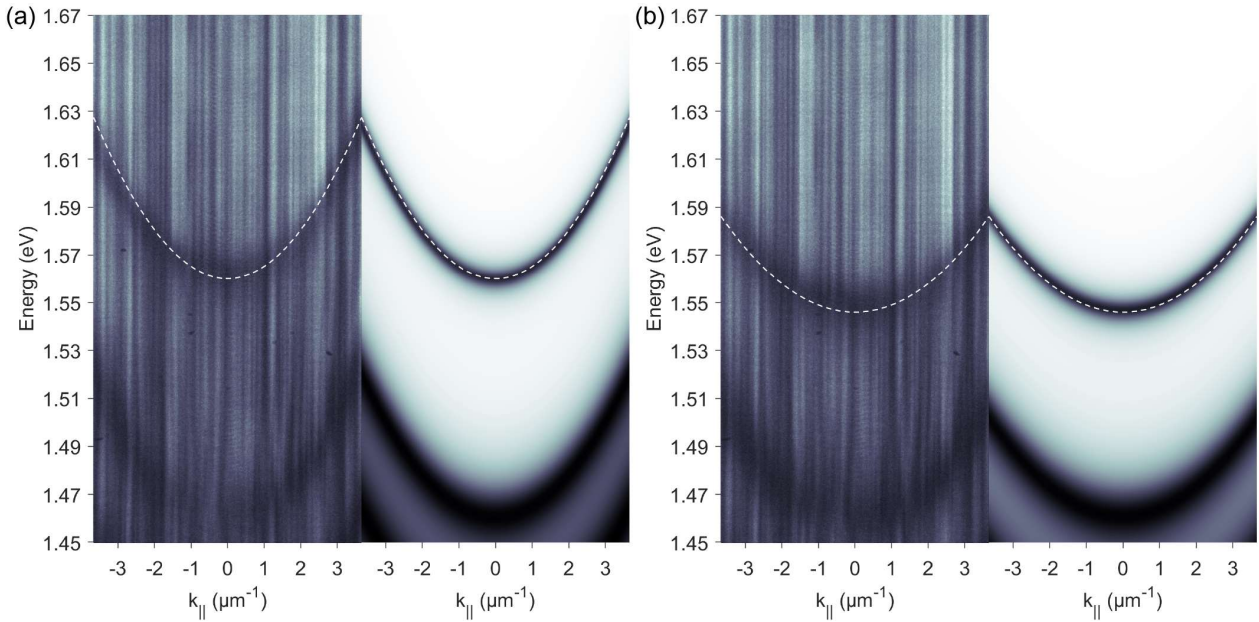

**Fig S4. Reflectance of ReS₂ microcavity taken at room temperature,** showing the dispersions of the bare cavity for $\widehat{d_1}$ (left, blue dashed line) and $\widehat{d_2}$ (right, white dashed line) polarized incident white light, with sample orientation α = 60°.

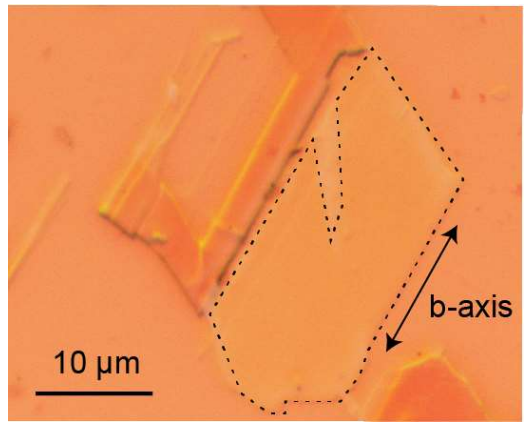

**Fig S5.** Optical microscope image of the ReS$_2$ sample



\end{document}